\documentclass[pra,preprint,showpacs]{revtex4}
\usepackage[centertags]{amsmath}
\usepackage{amsfonts}
\usepackage{amssymb}
\usepackage{amsthm} 
\usepackage{newlfont}
\usepackage{epsfig}
\usepackage{amscd}
\usepackage{graphicx}
\usepackage{epsfig}
\usepackage{footnote}
\usepackage{lipsum}
\usepackage{color}
\usepackage{xcolor}
\usepackage{graphicx}
\usepackage{multirow}
\usepackage{subfig}
\usepackage{amscd}

\newcommand{\beq}{\begin{equation}}
\newcommand{\eeq}{\end{equation}}
\newcommand{\ba}{\begin{array}}
\newcommand{\ea}{\end{array}}
\newcommand{\bea}{\begin{eqnarray}}
\newcommand{\eea}{\end{eqnarray}}

\begin{document}

\begin{center}
{\large \sc \bf {Simulation of three-spin evolution under XX Hamiltonian on quantum processor of  IBM-Quantum Experience}
}

\vskip 15pt

{\large 
S.I.Doronin, E.B.Fel'dman, and A.I.Zenchuk 
}

\vskip 8pt

{\it 
Institute of Problems of Chemical Physics RAS,
Chernogolovka, Moscow reg., 142432, Russia}

\end{center}

\begin{abstract}
We simulate the evolution of three-node spin chain on the quantum processor of IBM Quantum Experience using the diagonalization of $XX$-Hamiltonian and representing the evolution operator in terms of CNOT operations and one-qubit rotations.  We study the single excitation transfer from the first to the third node and show the significant difference between calculated and theoretical values of state transfer  probability. Then we propose a method reducing this difference by applying the two-parameter transformation including the shift and scale of the calculated probabilities. { We demonstrate the universality of this transformation inside of the class of three-node evolutionary systems governed by the $XX$-Hamiltonian.}

\end{abstract}

\noindent
{\it Corresponding author:} \\Alexander I. Zenchuk, 
e-mail: zenchuk@itp.ac.ru\\
ORCID: 0000-0003-3135-156X

\maketitle

\section{Introduction}
\label{Section:Introduction}

Development of quantum information is stimulated  by amazing advantages of quantum devices in comparison with their classical counterparts. In particular,  many well-known quantum algorithms {have appeared} \cite{NCh}.  Although they are written  for an ideal quantum processor,  these algorithms present a constitutional part of general progress in quantum computation. Among others we mention  Shor algorithm  \cite{Shor} for factorizing integers and   Deutsch-Jozsa   algorithm \cite{DJ} demonstrating quantum  speedup. There are also algorithms  allowing to perform set of algebraic operations \cite{ZZRF}. Most famous algorithm of this kind was proposed by  A.Harrow, A.Hassidim and S.Lloyd (HHL algorithm) \cite{HHL,CJS,BWPRWL} for solving systems of linear equations. This algorithm  includes 
the algorithm of  Hamiltonian simulation \cite{BACS,Ch} and phase estimation \cite{CEMM,LP}
based on the quantum Fourier transform \cite{NCh,GN}. { An essential part of those algorithm is a preparing particular input states which is a particular problem requiring  a special approach \cite{A,SLRV}. }

However, the implementation of those algorithms on the  real quantum processor faces large problems because of calculation errors. The fact is that 
contemporary quantum processors are far from the ideal ones. Operation of these processors unavoidably generates  quantum noise which creates a serious obstacle for nowaday application of  quantum algorithms in practice. It is known that  { quantum noise is mostly generated by}  two-qubit operations, such as CNOT { which is widely used in all algorithms to entangle different qubits.} Therefore the problem to compensate the effect of  quantum noise  is of a principal meaning. We study this problem simulating the quantum state evolution governed by the $XX$-Hamiltonian.

To simulate the spin-chain evolution under certain Hamiltonian on a quantum processor one can  formally appeal to  the Trotter method \cite{NCh,GGHCACCS,CHICCS}. This method was intensively studied and considerable  results were obtained 
\cite{ZRPL,CHICCS}. However, Trotterization involves numerous two-qubit operations which result in large calculation errors. 

{Therefore the alternative methods for simulating spin-evolution are of interest \cite{ZKEPL}. 
 As a simple example of such alternative, the evolution operator  for a two-qubit system governed by the $XX$-Hamiltonian was represented in terms of CNOT operations and one-qubit rotations in Refs.\cite{VD, GGHCACCS}.  

We develop a method of simulating the spin-evolution based on   diagonalization of the Hamiltonian, so that the time-dependent part 
of evolution operator becomes diagonal.}
 The obvious disadvantage of this method is that it is not completely quantum one but involves the classical operation of Hamiltonian diagonalization at the first stage of calculation. However, the simplicity of realization of the diagonal evolution in comparison with the nondiagonal one is very promising.  

Here we consider the evolution of short two- and three-qubit spin systems based on the diagonalization of the $XX$-Hamiltonian
\begin{eqnarray}
H=U \Lambda U^+,
\end{eqnarray}
where $\Lambda$ is the matrix of eigenvalues and $U$ is the matrix of eigenvectors of the Hamiltonian. 
Therefore the evolution operator $V$ can be written as
\begin{eqnarray}\label{evH}
V(t)=  e^{-i  H t}  = U e^{-i  \Lambda t} U^+.
\end{eqnarray}
Thus, the evolution is hidden into the diagonal operator $e^{-i  \Lambda t}$. 

Below we propose the scheme including the CNOT operations and one-qubit rotations which realize both the unitary 
transformation $U$  and the diagonal evolution $e^{-i  \Lambda t}$ in (\ref{evH}) and which can be executed on a quantum processor. Then we simulate  the propagation of the one-qubit excitation from the first to the third node and compare the  probability of the excited state transfer calculated on a quantum processor with  the theoretical value of this probability. As expected, the deviation is significant. But using the special transformation of the calculated  probability we can significantly improve the result. Such transformation is a principal subject of our paper. 

The paper is organized as follows. In Sec.\ref{Section:2spins}, we consider the two-spin chain governed by the $XX$-Hamiltonian and  give diagonal representation of the evolution operator in terms of CNOT operations and one-qubit rotations. Then, in Sec.\ref{Section:3spins}, we give detailed description of the three-spin alternating chain governed by the $XX$-Hamiltonian under approximation of nearest-neighbor interactions. For different values of alternation parameter, we consider the probability of the excited state transfer from the first to the third spin and compare the 
theoretically predicted value of this probability with the value found on the quantum processor. We also consider a two-parametric transformation reducing the difference between the two above values. We demonstrate certain  universality of that transformation  by applying it to the three-node alternating chain governed by the $XX$-Hamiltonian with all-node interactions. Conclusions are  given in Sec.\ref{Section:Conclusion}.

\section{Two-spin chain}
\label{Section:2spins}
We start with  the  evolution of a dimer under the  
$XX$ Hamiltonian
\begin{eqnarray}\label{2HH}
H_2=D(I_{x1} I_{x2} + I_{y1}I_{y2})
\end{eqnarray}
(where $D$ is a coupling constant, $I_{xi}$ and  $I_{yi}$ are the $x$- and $y$-projections of the angular momentum of the $i$th spin) and simulate it  on the quantum processor. For this purpose we diagonalise  Hamiltonian (\ref{2HH}):
\begin{eqnarray}
H_2=U_{12} \Lambda_{12} U^+_{12},
\end{eqnarray}
where 
$U_{12}$ and $\Lambda_{12}$ are, respectively, the matrices of  eigenvectors and eigenvalues,  
\begin{eqnarray}\label{U12}
U_{12}=\left(
\begin{array}{cccc}
1&0&0&0\cr
0&\frac{1}{\sqrt{2}}&\frac{1}{\sqrt{2}}& 0\cr
0&\frac{1}{\sqrt{2}}&-\frac{1}{\sqrt{2}}& 0\cr
0&0&0&1
\end{array}
\right),\;\;\Lambda_{12}=D \tilde \Lambda_{12},\;\; \tilde \Lambda_{12} =  {\mbox{diag}}(0,2,-2,0).
\end{eqnarray}
Evolution operator (\ref{evH})  in the diagonalised form  reads
\begin{eqnarray}\label{evH2}
V(t)  = U_{12} e^{-i  \tilde \Lambda_{12} \tau} U^+_{12} ,\;\;\tau=D t,
\end{eqnarray}
where $\tau$ is the dimensionless time.

\begin{figure*}[!]
\epsfig{file=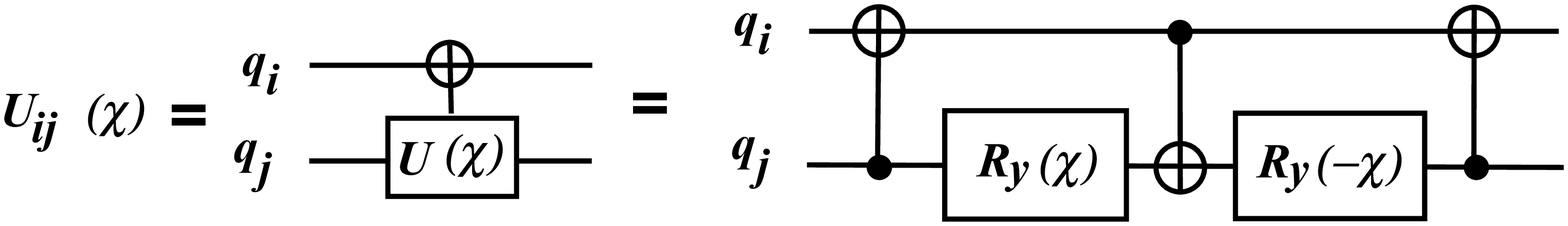,
  scale=0.4
   ,angle=0
} 
\caption{Simplest one-parameter two-qubit operator $U_{ij}$ conserving the excitation number in a system.
}
  \label{Fig:U} 
\end{figure*}

It is remarkable that the operator  $U_{12}$
 can be written in terms of CNOT operations and one-qubit $y$-rotations as follows:
\begin{eqnarray}\label{U}
U_{12}=C_{21} R_{y2}(\frac{\pi}{4}) C_{12} R_{y2}(-\frac{\pi}{4}) C_{21},
\end{eqnarray}
where $R_{yj}$ is the $y$-rotation of the $j$th spin over the angle $\phi$:
\begin{eqnarray}
R_{yj}(\phi)=e^{-i I_{yj} \phi},
\end{eqnarray}
while $C_{ij}$ and $C_{ji}$ written in the basis
\begin{eqnarray}
(00), (0j),(i0),(ij)
\end{eqnarray}
read respectively
\begin{eqnarray}
C_{ij}=\left(
\begin{array}{cccc}
1&0&0&0\cr
0&1&0&0\cr
0&0&0&1\cr
0&0&1&0
\end{array}
\right),\;\;C_{ji}=\left(
\begin{array}{cccc}
1&0&0&0\cr
0&0&0&1\cr
0&0&1&0\cr
0&1&0&0
\end{array}
\right).
\end{eqnarray}
The operator $U_{12}$  is a particular case  of the operator 
\begin{eqnarray}\label{Uchi}
U_{ij}(\chi)=C_{ji} R_{yj}(\chi) C_{ij} R_{yj}(-\chi) C_{ji},
\end{eqnarray}
introduced in Ref.\cite{DFZ_2020} as a simplest two-qubit operator conserving the number of excited spins in a system (i.e., commuting with the $z$-projection of the total spin momentum $I_z$) and  representable  in terms of CNOT operations and one-qubit $y$-rotations, Fig.\ref{Fig:U}.  Namely this operator will be used as a structural block  for constructing the eigenvector matrix for the 3-qubit systems below.

In turn, the diagonal operator $e^{-i \tilde \Lambda_{12} \tau}$ can be written as a composition of two one-qubit $z$-rotations,
\begin{eqnarray}
&&
e^{-i \tilde \Lambda_{12}\tau} = R_{z1}(2\tau) R_{z2}(-2 \tau),\\
&&
R_{zi}(\phi)=e^{-i I_{zi} \phi},
\end{eqnarray}
which can be simply simulated on a quantum processor. 

The two-qubit systems have been studied, for instance, in \cite{GGHCACCS,VD}, therefore  we do not consider them   here. { Notice also that our two-qubit evolution operator (\ref{evH2}) includes 6 CNOT operations which is rather big number of two-qubit operations. But the constructed  unitary block $U_{ij}$    (\ref{Uchi}) preserves the excitation number and can be used in systems of higher dimension, which is shown below in 
Sec.\ref{Section:3spins} for a three-spin system.} 

\section{Three-spin chain}
\label{Section:3spins}
We consider the evolution of three-spin system under the $XX$-Hamiltonian using both approximation of nearest neighbor interactions and all spin interactions.
 { We  show difference and similarity  in simulation schemes} for each of these cases on a quantum computer.

\subsection{Nearest neighbor approximation}
\label{Section:NNA}
First  we consider the three-node chain governed by the $XX$-Hamiltonian using  the nearest neighbor approximation with two different coupling constants  (an alternating chain), 
\begin{eqnarray}\label{HH0}
H_3=D(I_{x1} I_{x2} + I_{y1}I_{y2} + d (I_{x2} I_{x3} + I_{y2}I_{y3})).
\end{eqnarray}
where $d$ is the alternation parameter ($d=1$ for the homogeneous chain).
We can diagonalise this Hamiltonian,
\begin{eqnarray}
\label{HH}
&&
H_3=U_{123} \Lambda_{123} U^+_{123}.
\end{eqnarray}
Then the evolution operator (\ref{evH}) reads 
\begin{eqnarray}\label{ev3}
V(t)= U_{123} e^{-i\tilde{\Lambda}_{123}} U^+_{123},\;\;\tau = D t.
\end{eqnarray}
Here $U_{123}$ and $\Lambda_{123}$ are the matrices of eigenvectors and eigenvalues respectively. We do not give the explicit form for $U_{123}$,  but represent 
 the matrix  $U_{123}$  in terms of CNOT operations and one-qubit $y$-rotations as follows:
\begin{eqnarray}\label{U123}
U_{123}= SWAP_{23}V^{(1)}_{123} U_{12}  U_{23} V^{(2)}_{123}SWAP_{23}.
\end{eqnarray}
{ Here  $SWAP_{23} = C_{23} C_{32} C_{23}$, the operators $V^{(i)}_{123}$, $i=1,2$, are the diagonal, and $V^{(1)}_{123}$ is representable in terms of CNOT operations and $z$-rotations,
\begin{eqnarray}
V^{(1)}_{123}= R_{z1}(\frac{\pi}{2}) C_{32} R_{z2}(\frac{\pi}{2}) C_{32}. \;\;
\end{eqnarray}
while the operator $V^{(2)}_{123}$,
\begin{eqnarray}
V^{(2)}_{123}={\mbox{diag}}(-i,1,1,i,1,i,i,1),
\end{eqnarray}
commutes with $SWAP_{23}$, 
\begin{eqnarray}\label{SWAPc}
[V^{(2)}_{123}, SWAP_{23}]=0,
\end{eqnarray}
and therefore  disappears from Hamiltonian (\ref{HH}).}
Each of the operators $U_{12}$ and $U_{23}$  in (\ref{U123}) is of form (\ref{Uchi}) with  a particular choice of the parameter $\chi$. 
In the  operator $U_{23}$, we take  $\chi=\frac{\pi}{4}$, so that
\begin{eqnarray}
U_{23}=C_{32} R_{y3}(\frac{\pi}{4}) C_{23} R_{y3}(-\frac{\pi}{4}) C_{32}.
\end{eqnarray}
Written in the matrix form, this operator coincides with the operator $U_{12}$ (\ref{U12})  in  Sec. \ref{Section:2spins}.
%
The parameter $\chi$ in the operator $U_{12}$ in (\ref{U123}) is defined by the alternation parameter $d$,
\begin{eqnarray}
&&U_{12}=C_{21} R_{y2}(\chi) C_{12} R_{y2}(-\chi) C_{21},\\\nonumber
&&
U_{12}=\left(
\begin{array}{cccc}
1&0&0&0\cr
0&\sin \chi &\cos\chi& 0\cr
0&\cos\chi&-\sin\chi& 0\cr
0&0&0&1
\end{array}
\right),
\;\; d =\tan\,\chi.
\end{eqnarray}
{Since $d>0$, we consider $0<\chi<\frac{\pi}{2}$. }
In a particular case of  homogeneous chain  $d=1$ and $\chi=\frac{\pi}{4}$. Therefore the operators 
$U_{23}$ and $U_{12}$ become equivalent two-qubit operators, each coincides with the operator $U_{12}$ in Sec.  \ref{Section:2spins}, but is applied to different pairs of qubits.

\begin{figure*}[!]
\epsfig{file=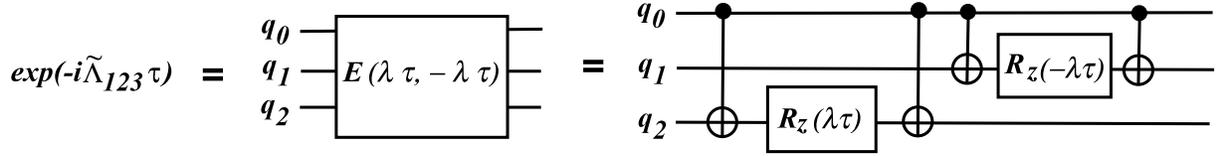,
  scale=0.4
   ,angle=0
} 
\caption{Simulation of diagonal part $e^{-i\tilde\Lambda \tau}$ of the three-qubit  evolution operator
in terms of CNOT operations and one-qubit $z$-rotations.
}
  \label{Fig:Lam} 
\end{figure*}

Finally, the eigenvalue matrix $\Lambda_{123}$ reads
\begin{eqnarray}
\Lambda_{123} = D \tilde\Lambda_{123},\;\; \tilde\Lambda_{123}={\mbox{diag}}(0,-\lambda,\lambda,0,0,\lambda,-\lambda,0),\;\;
\lambda = \frac{\sqrt{1+d^2}}{2} = \frac{1}{2 \cos \,\chi}.
\end{eqnarray}
Then the diagonal part of the evolution operator $e^{-i \tilde{\Lambda}_{123} \tau}$ can be represented in terms of CNOT operations  and $z$-rotations as follows, see Fig.\ref{Fig:Lam}:
\begin{eqnarray}\label{ev}
e^{-i \tilde \Lambda_{123} \tau} = C_{12} R_{z2}(-\lambda \tau) C_{12}  C_{13} R_{z3}(\lambda\tau) C_{13}.
\end{eqnarray}

\subsubsection{Probability of excited state transfer: theoretical results versus simulation on quantum processor}

\begin{figure*}[!]
\epsfig{file=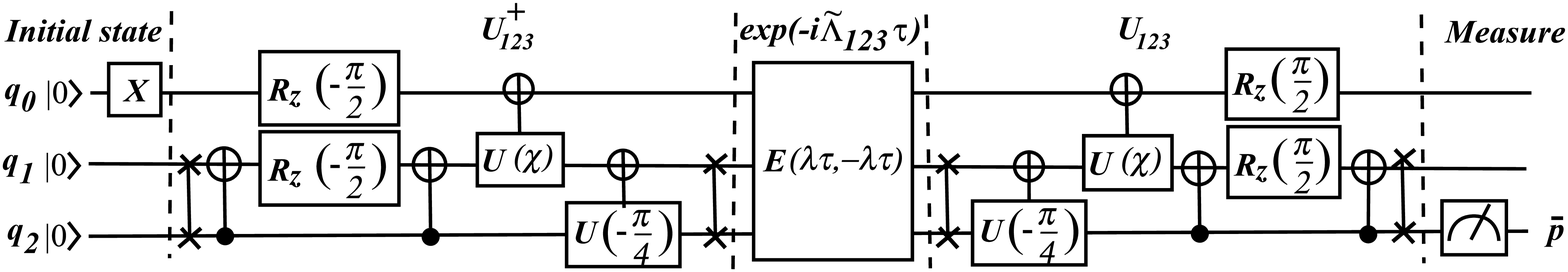 ,
  scale=0.35
   ,angle=0
} 
\caption{Simulation  of the excited state evolution along the three-spin chain on the quantum processor. The probability $\bar{p}$ (\ref{theor1}) of the excitation transfer is an output. Here $\chi=\tan^{-1} d$.
}
  \label{Fig:S} 
\end{figure*}

We study the probability of the excited state  transfer from the first to the third spin, i.e., the initial 
state of the three-qubit chain reads
$|100\rangle$. This state governed by the Hamiltonian (\ref{HH0}), (\ref{HH}) evolves as follows:
\begin{eqnarray}\label{psi}\label{theor1}
|\psi(\tau)\rangle& =& U_{123}e^{-i \tilde \Lambda_{123} \tau} U^+_{123}|100\rangle=
\cos \chi \sin \chi(\cos(\lambda \tau)-1){|001\rangle -} \\\nonumber
&&{i \cos \chi \sin (\lambda \tau )
|010\rangle + (\cos^2 \chi (\cos( \lambda \tau)-1) +1) |100\rangle.}
\end{eqnarray}
This evolution can be simulated on a quantum processor with the scheme  in Fig.\ref{Fig:S}. 
The probability of the excited state transfer to the third spin is following:
\begin{eqnarray}\label{theor0}
p= \big|\langle 001|\psi(\tau)\rangle\big|^2 ={ \frac{1}{4} \sin^2(2\chi)(\cos(\lambda \tau)-1)^2 .}
\end{eqnarray} 
Namely this probability is measured at the third qubit $q_2$ as an output, as  shown in Fig.\ref{Fig:S}. 
{ We implement our quantum protocol on the 5-qubit quantum processor 
{\it ibmq\_lima} of IBM Quantum Experience using $2^{13}$ shots and averaging over 5 runs.}

{ Having probability obtained both with  analytical formulas (\ref{theor1}) and (\ref{theor0}),  and as the result of  calculation on the quantum processor, we discuss a method for reducing the discrepancy between those two values.} 
We emphasize that the scheme in Fig.\ref{Fig:S} serves to describe the evolution of any initial state  at the input site of the scheme. Of course, it can be simplified for transfer just single excited state. We do not consider such simplification.

\begin{figure*}[!]
\epsfig{file=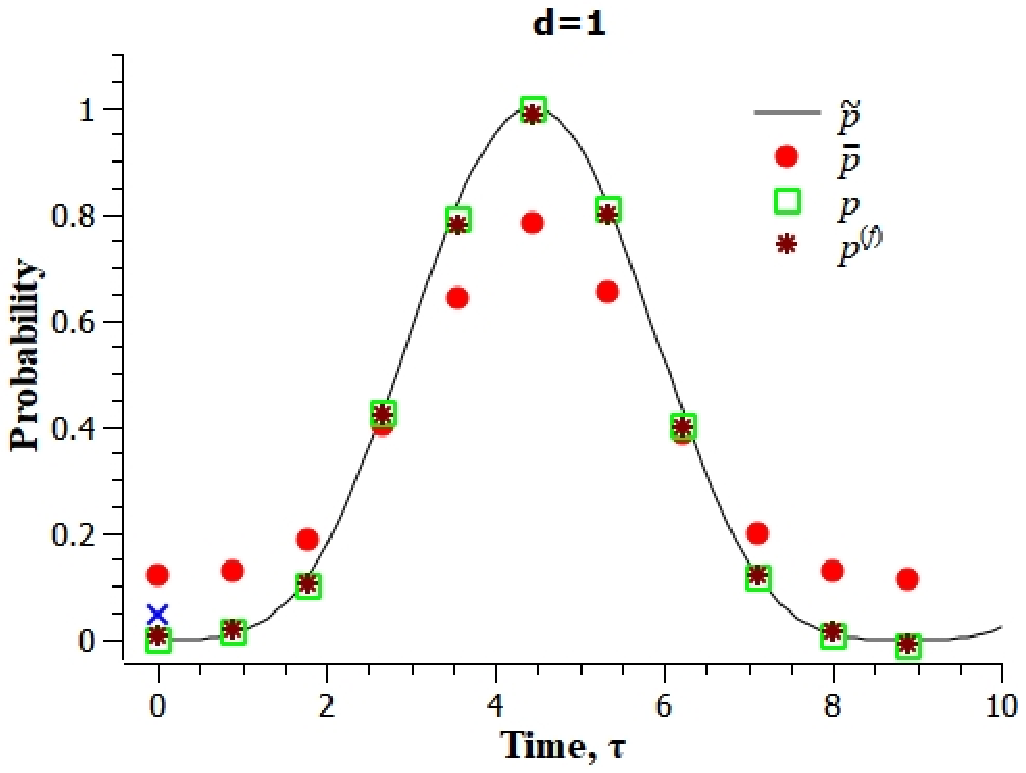,
  scale=0.7
   ,angle=0
}
\epsfig{file=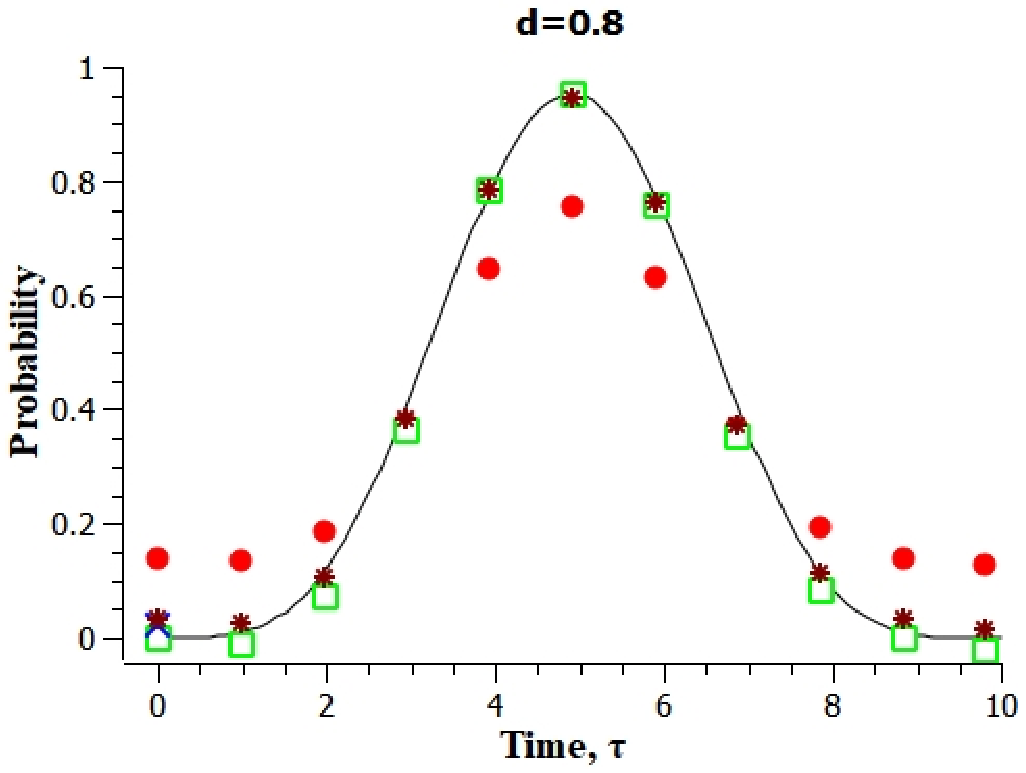,
  scale=0.7
   ,angle=0
} 
\epsfig{file=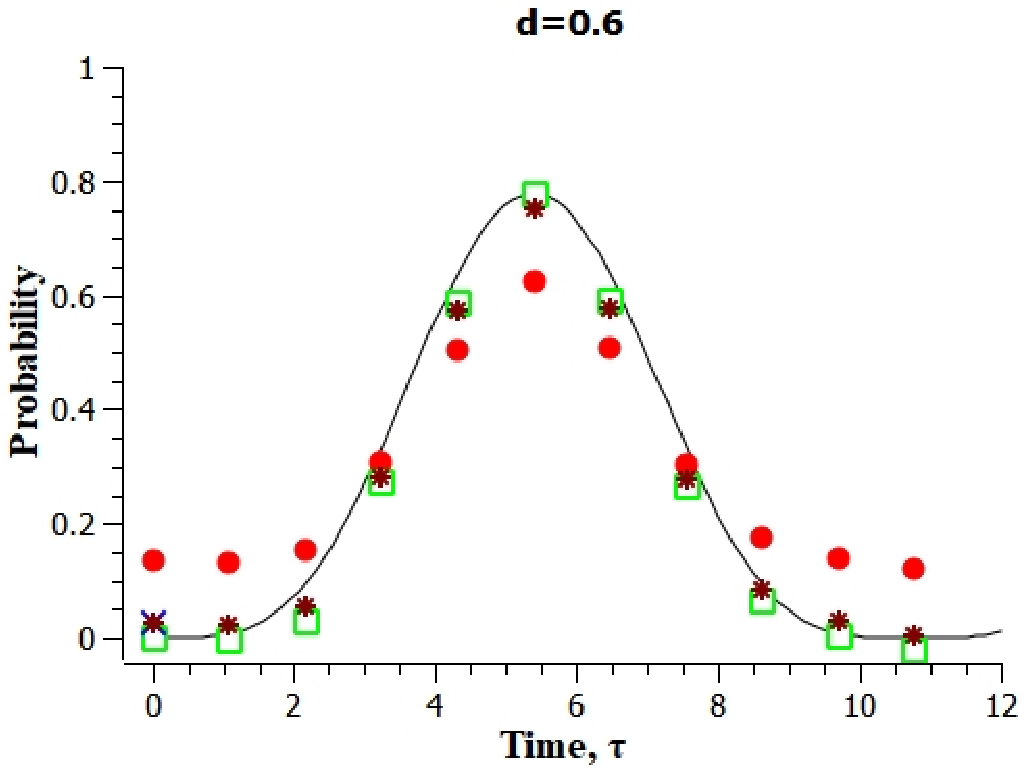,
  scale=0.7
   ,angle=0
} 
\epsfig{file=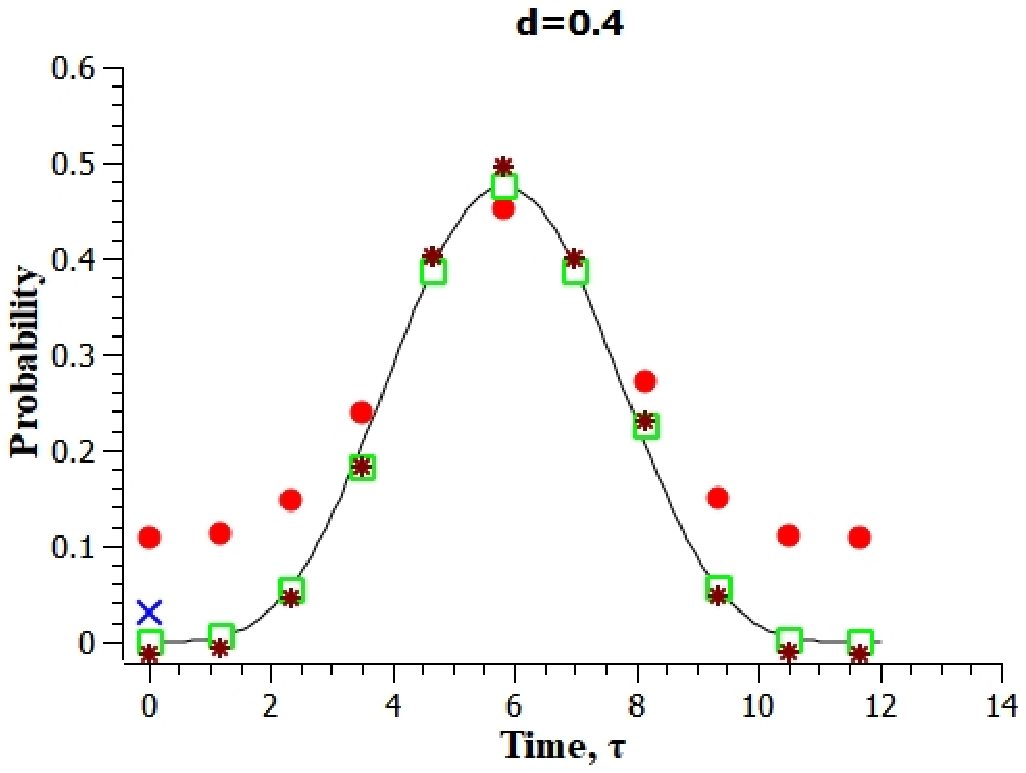,
  scale=0.7
   ,angle=0
} 
\epsfig{file=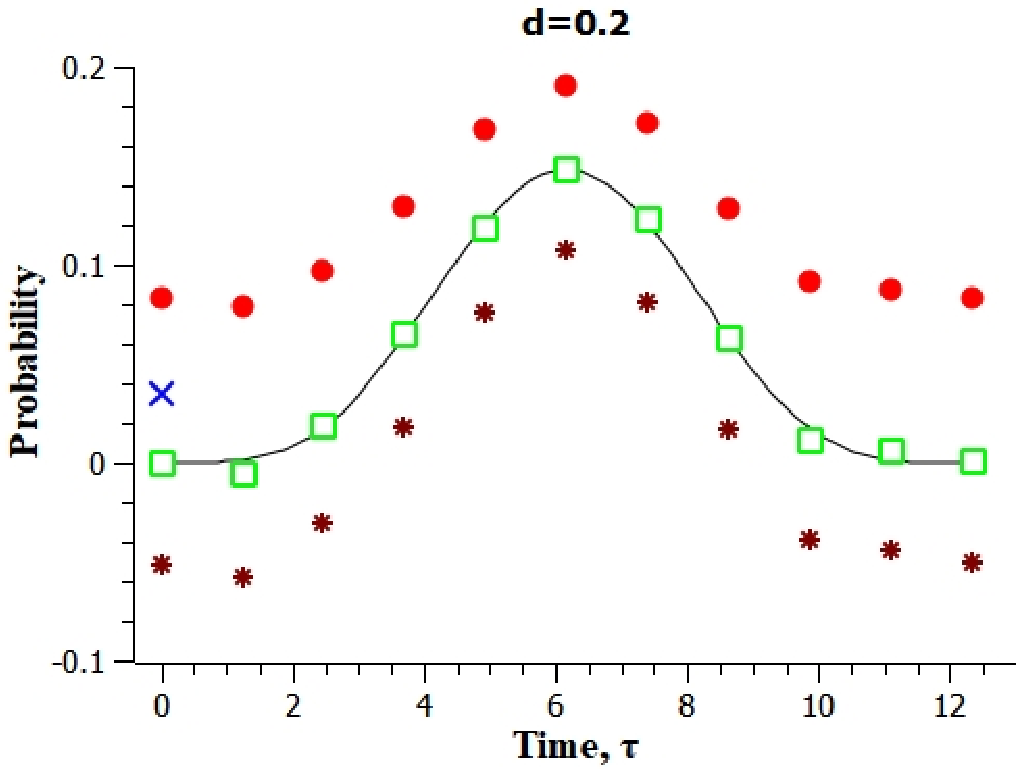,
  scale=0.7
   ,angle=0
} 
\caption{ Probability of the excited state transfer from the 1st to the 3rd spin  of the alternating chain governed by  the $XX$-Hamiltonian, approximation of nearest-node interactions. Different plates correspond to different values of the alternation parameter $d$, $d=1$ corresponds to the homogeneous chain.
{  
We present the theoretically calculated  probability  $p$ (solid lines), probability calculated   on the quantum processor $\tilde p$(circles),  corrected probabilities $\bar{p}$  (squares) and $p^{(f)}$  (stars), { cross means probability measured at the quantum processor at $\tau=0$, see text}. 
}}
  \label{Fig:QE1} 
\end{figure*}

Thus, the results of calculation on quantum processor of IBM QE exhibit  significant difference in comparison with the theoretical results, as  shown in Fig.\ref{Fig:QE1}, where theoretically calculated $p$ is represented by the solid line, while the probabilities   calculated on the quantum processor  are represented by circles for a set of values of the alternation parameter $d$:
$d=1,\;0.8,\;0.6,\;0.4,\;0.2$.
{On this figure, the largest difference is revealed for small values of $p$ ($\tilde p>p$) as well as for large values of $p$ ($\tilde p<p$), whereas $\tilde p \approx p$ for intermediate values. This observation can be clarified as follows. If  the theoretical  probability to detect the excitation at a given qubit is negligible ($p\to 0$), then measured probability is increased due to the  noise  of quantum gates. On the contrary, if the theoretical  probability is significant ($p\sim 1$), then  the same noise decreases  the measured value of this probability. Thus, there are two  counteracting  noise effects: increasing small probability of excitation registration and  decreasing large value of this probability. As the result of this counteracting, the intermediate values of probability remains almost correct. If that is the basic  reason of the above observed difference between $p$ and $\tilde p$, then we can try to eliminate (or at least reduce) it by the regular method. One of such methods might be analogous to the geometrical  homotetic transformation. For that, one needs to fix the homotetic center (i.e., such value $p_0$ that $p_0=\tilde p_0$) and calculate the homotetic coefficient. This approach requires additional detailed exploration of the relations between the theoretically predicted and measured probabilities so that we postpone it. Instead, we suggest a simpler method which nevertheless gives a quite acceptable result. }

It is interesting to note that the  difference between the calculated value $\tilde{p}(\tau)$ and  theoretical one
$p(\tau)$ can be reduced  by the combination of two operations: 
the shift over the value $s$ and  scale by the factor $k$ defined as follows:
\begin{eqnarray}\label{sk}
&&
s=\tilde{p}(\varepsilon)-p(\varepsilon),\;\; \varepsilon  = 0.001 \ll 1,\\\nonumber
&&
k = p_{max}/\tilde{p}_{max}. 
\end{eqnarray}
{ Here $\tau_{max}$ is a time instant 
of maximum of $p(\tau)$: $p_{max}=p(\tau_{max})$ ,  and  $\tilde{p}_{max}=\tilde{p}(\tau_{max})$. Thus, we obtain the corrected function $\bar{p}$,
\begin{eqnarray}\label{barp}
\bar{p} = k (\tilde{p} - s),
\end{eqnarray}
shown in Fig.\ref{Fig:QE1} by squares. We see that most squares belong to the solid curve representing the theoretical 
probability of the excited  state transfer.}

\begin{figure*}[!]
\epsfig{file=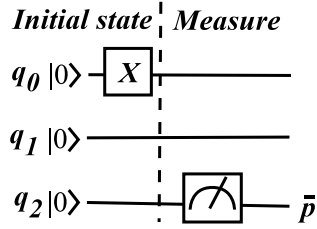,
  scale=0.4
   ,angle=0
}
\caption{{Scheme in Fig.\ref{Fig:S} trivialized at $\tau=0$. }  
}
\label{Fig:S2} 
\end{figure*}

{
We shall clarify the reason of setting $\tau=\varepsilon$ instead of $\tau=0$ in the definition of the shift $s$, Eq.  (\ref{sk}). 
The matter is that the quantum processor can automatically simplify the simulation scheme canceling the combinations of the form  $AA^+$, where $A$ is a unitary transformation. Therefore, if $\tau=0$, then the evolution operator $V$ becomes the identity $E$:  $V(0)= U_{123} U^+_{123}=E$. Therefore the scheme in (\ref{Fig:S}) is drastically simplified { up to} the scheme in Fig.\ref{Fig:S2}. Thus, setting $\tau=0$ we examine a trivial scheme and the measured probability is different, see crosses in Fig.\ref{Fig:QE1}. The minor deviation $\varepsilon$  from $\tau=0$ turns on all operators in 
Fig.\ref{Fig:S} .} 
The shifts $s$ and scale factors $k$ for different values of the alternation parameter $d$ are collected in Table \ref{Table:1}.
\begin{table}
\begin{tabular}{|c|ccccc|c|}
\hline
$d$&1&0.8&0.6&0.4&0.2&mean\cr
\hline
$s$&0.1211&0.1400&0.1352&0.1091&0.0827&0.1176\cr
$k$&1.5145&1.5475&1.5924&1.3873&1.3761&1.4836\cr
\hline
\end{tabular}
\caption{Shifts $s$ and scale factors $k$ for set of values of the alternation parameter $d$. The last column presents the mean values $\bar{s}$ ans $\bar{k}$.}
\label{Table:1}
\end{table}
Remember, that the {analogous transformation} of the computed probabilities with the purpose to fit the theoretical results was  proposed in \cite{ZRPL} to correct errors of  implementation of the Trotter method  of Hamiltonian simulation on a quantum processor.


Table \ref{Table:1} shows that the spread of shifts $s$ and scales $k$ found for different experiments is not significant,  especially for $d>0.2$. This motivates us to introduced the mean values $\bar{s}$ and $\bar{k}$ for these shifts and scales (arithmetical averages),
\begin{eqnarray}
\bar{s}=0.1176,\;\;\bar{k}=1.4836 ,
\end{eqnarray}
and use this values to correct the probabilities calculated on quantum processor in all experiments with the scheme in Fig.\ref{Fig:S}.  
Thus, we propose to use the corrected probability $p^{(f)}$,
\begin{eqnarray}\label{pf}
p^{(f)} = \bar{k} (\tilde{p} - \bar{s}),
\end{eqnarray}
instead of $\bar{p}$ given in (\ref{barp}).
In fact, the  probability $p^{(f)}$ shown in Fig.\ref{Fig:QE1} by stars only slightly differs from the probability 
$\bar{p}$ shown by squares,  except the case $d=0.2$.  { The last fact is in accordance with the observation made in \cite{DFZ_2020}, where the corrected quantity well agrees with the theoretical value   for the case when  the measured probability exceeds $\sim 0.2$. }

Below we use  the {calculated} $\bar{s}$ and $\bar{k}$ to find corrected probabilities of the excited state transfer   along the  three-node chains with {\it all-node interactions} governed  by the XX-Hamiltonian. 

\subsection{$XX$-Hamiltonian with all-node interactions}
The 3-qubit  
$XX$-Hamiltonian  involving interactions among  all-nodes reads:
\begin{eqnarray}\label{HH2}
&&
H_3=D_{12}(I_{x1} I_{x2} + I_{y1}I_{y2}) + D_{23} (I_{x2} I_{x3} + I_{y2}I_{y3}) + 
D_{13} (I_{x1} I_{x3} + I_{y1}I_{y3}),\\\nonumber
&&D_{ij} = \frac{\gamma^2 \hbar}{r_{ij}}^3,
\end{eqnarray}
where $\gamma$ is the gyromagnetic ratio, $\hbar$ is the Planck constant, $r_{ij}$ is the distance between the $i$th and $j$th nodes. 
The diagonalization of this Hamiltonian is given by formula (\ref{HH}) with different $U_{123}$ and $\lambda_{123}$.
Unlike the case of nearest neighbor approximation, the representation of the evolution of the  homogeneous spin chain in terms of CNOT operations and one-qubit rotations can not be considered as a limit $d\to 1$ { of the similar representation for} the alternating chain. 
It seamed out that $d\neq 1$ requires additional operator $U$ of form (\ref{Uchi}) and consequently includes more CNOT  operators. 
Therefore, we consider the homogeneous and alternating chains  separately. 

\subsubsection{Homogeneous chain}
For the homogeneous chain 
we set 
$D_{12}= D_{23}=D$ and use the  dimensionless time 
$\tau = D t$.
We have 
\begin{eqnarray}\label{2U123}
U_{123}&=& SWAP_{23}V^{(1)}_{123} U_{12}  U_{23}  V^{(2)}_{123} SWAP_{23},\\\nonumber
&&V^{(1)}_{123} = R_{z1}(\frac{\pi}{2}) C_{32} R_{z2}(\frac{\pi}{2})C_{32} ,\\\nonumber
&& V^{(2)}_{123} ={\mbox{diag}}(-i,1,1,i,1,i,i,1),
\end{eqnarray}
where $V^{(2)}_{123}$ commutes with $SWAP_{23}$ and 
therefore disappears from Hamiltonian (\ref{HH2}).
Next, we find
\begin{eqnarray}
&&U_{12}= C_{21} R_{y2}(\chi_1) C_{12} R_{y2}(-\chi_1)C_{21},\\\nonumber
&&U_{23}= C_{32} R_{y3}(\chi_2) C_{23} R_{y3}(-\chi_2)C_{32},
\end{eqnarray}
where
\begin{eqnarray}
\chi_1 = \frac{\pi}{4}, \;\chi_2 = \arccos \left(\frac{16}{\sqrt{513-3\sqrt{57}}}\right)\approx 0.7633.
\end{eqnarray}
Finally, 
\begin{eqnarray}
&&\Lambda_{123}= D_{12} \;\;{\mbox{diag}} (0, \lambda_1, \lambda_2, -\lambda_3, -\lambda_3, \lambda_2, \lambda_1, 0),\\\nonumber
&&
\lambda_1= \frac{1}{32}(1-3\sqrt{57}), \; \lambda_2=\frac{1}{32}(1+3\sqrt{57}),\; \lambda_3= \frac{1}{16},
\end{eqnarray}
so that the diagonal part of the evolution operator reads:
\begin{eqnarray}\label{Lambda2}
e^{-i \tilde \Lambda_{123} \tau} =
C_{21} R_{z1} (\lambda_1 \tau) C_{21} C_{32} R_{z2} (-\lambda_3 \tau) C_{32} C_{13} R_{z3}(\lambda_2\tau) C_{13}.
\end{eqnarray}
The evolution  of the first excited  spin is described by the formula
\begin{eqnarray}\label{theor2}
|\psi(t)\rangle&=& U_{123}e^{-i \tilde \Lambda_{123} \tau} U^+_{123}|100\rangle = 
\frac{1}{2} \Big(\sin^2 \chi_2 e^{-i\lambda_1 \tau}+ \cos^2 \chi_2 e^{-i\lambda_2 \tau} -  e^{i\lambda_3\tau}\Big)|001\rangle+\\\nonumber
&&
\gamma_1(\tau)|010\rangle +\gamma_2(\tau) |100\rangle,
\end{eqnarray}
where we do not give explicite form for $\gamma_i$, $i=1,2$. 
Then the probability of the excited state  transfer is defined  by Eq.(\ref{theor0}).

We notice that the representation of the evolution operator in terms of CNOT operations and one-spin rotations coincides with such representation of the evolution operator {for} $XX$-Hamiltonian under the approximation of  nearest-neighbor interactions  described in Sec.\ref{Section:NNA}. Therefore, the corrected probability $p^{(f)}$ (\ref{pf}) can be found with the same mean values $\bar{s}$ and $\bar{k}$, see Table \ref{Table:1}.  
The scheme of implementation of such state transfer on the quantum processor coincides with one given in Fig.\ref{Fig:S} with the only difference in the $\tau$-dependent operator $\exp({-i\tilde{\Lambda}_{123}\tau})$ which now is given in Eq.(\ref{Lambda2}) and includes one more rotation and two additional CNOT operators in comparison with Eq.(\ref{ev}). 

\begin{figure*}[!]
\epsfig{file=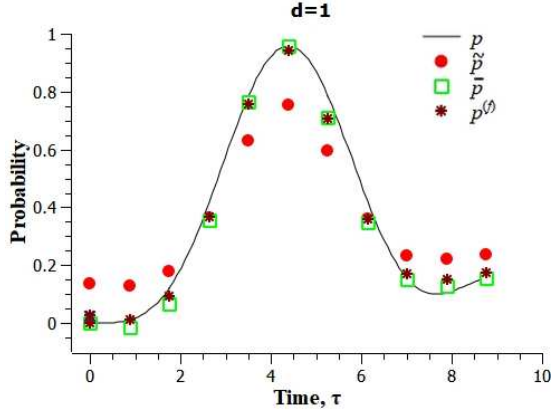,
  scale=0.7
   ,angle=0
}
\caption{ Probability of the excited state transfer from the 1st to the 3rd spin  of the homogeneous chain governed by  the $XX$-Hamiltonian, interactions among all nodes are included.
{  
We present the theoretically calculated  probability  $p$ (solid lines), probability calculated   on the quantum processor $\tilde p$(circles),  corrected probabilities $\bar{p}$  (squares) and $p^{(f)}$  (stars), { cross means probability measured at the quantum processor at $\tau=0$}. 
}
{To construct $\bar{p}$  we use $s=0.1357$ and $k= 1.5541$ in Eq.(\ref{sk}) }. 
}
  \label{Fig:QE2} 
\end{figure*}

In Fig.\ref{Fig:QE2}, we
see that the stars $p^{(f)}$ very closely approach the theoretical curve, similar to  Fig.\ref{Fig:QE1}. { This confirms the applicability of the corrected probability $p^{(f)}$ to this case of Hamiltonian.}

\subsubsection{Alternating chain with $d=1/2$.}
In this section we consider an example of state transfer, for which the scheme of its implementation on the quantum processor differs from one  considered in Fig.\ref{Fig:S}, but 
nevertheless the corrected probability $p^{(f)}$ with the parameter $\bar{s}$ and $\bar{k}$ from Table \ref{Table:1} also significantly improves the result.

We  use Hamiltonian (\ref{HH2}), dimensionless time $\tau = D_{12} t$ and alternation constant 
$d=\frac{D_{23}}{D_{12}}$.  {For certainty, we set   $d=1/2$.}
In Eq.(\ref{HH}) we have 
\begin{eqnarray}\label{3U123}
U_{123}= SWAP_{23}V^{(1)}_{123} U_{12} V^{(2)}_{123} U_{23} U_{31} V^{(3)}_{123}SWAP_{23}.
\end{eqnarray}
\begin{eqnarray}
&&V^{(1)}_{123} = R_{z1}(\frac{\pi}{2}) C_{23} R_{z3}(\frac{\pi}{2})C_{23} ,\\\nonumber
&&V^{(2)}_{123} = C_{12}R_{z2}(\pi) C_{12},\\\nonumber
&& V^{(3)}_{123} ={\mbox{diag}}(-1,-i,-i,1,i,-1,-1,-i).
\end{eqnarray}
Obviously, $V^{(3)}_{123}$ commutes with $SWAP_{23}$: $[V^{(3)}_{123}, SWAP_{23}]=0$.
Therefore this diagonal operator disappears from Hamiltonian (\ref{HH2}).
Next,
\begin{eqnarray}
&&U_{12}= C_{21} R_{y2}(\chi_1) C_{12} R_{y2}(-\chi_1)C_{21},\\\nonumber
&&U_{23}= C_{32} R_{y3}(\chi_2) C_{23} R_{y3}(-\chi_2)C_{32},\\\nonumber
&&U_{31}= C_{13} R_{y1}(\chi_3) C_{31} R_{y1}(-\chi_3)C_{13},
\end{eqnarray}
where
\begin{eqnarray}
\chi_1 = 3.5541, \;\chi_2 = 0.7712,\; \chi_3 = 1.6380.
\end{eqnarray}
Finally, 
\begin{eqnarray}\label{3L123}
&&\Lambda_{123}= D_{12} {\mbox{diag}} (0, -\lambda_1, \lambda_2, -\lambda_3, -\lambda_3, \lambda_2, -\lambda_1, 0),\\\nonumber
&&
\lambda_1= 0.5426, \; \lambda_2=0.5772,\; \lambda_3= 0.0346.
\end{eqnarray}
So that the diagonal part of the evolution operator reads:
\begin{eqnarray}\label{3ev}
e^{-i \tilde \Lambda_{123} \tau} =
C_{21} R_{z1} (-\lambda_1\tau) C_{21} C_{32} R_{z2} (-\lambda_3\tau) C_{32} C_{13} R_{z3}(\lambda_2\tau) C_{13}.
\end{eqnarray}
{and structurally coincides with Eq.(\ref{Lambda2}). }

The  evolution of the initial excited state reads as follows:
\begin{eqnarray}\label{theor3}
&&
|\psi(\tau)\rangle =  U_{123}e^{-i \tilde \Lambda_{123} \tau} U^+_{123}|100\rangle=\Big(\frac{e^{i \lambda_1\tau}}{2}
\sin 2\chi_1 \sin^2\chi_2
-\\\nonumber
&&
e^{-i \lambda_2\tau} (\cos\chi_1 \cos \chi_3 - \sin\chi_1 \cos\chi_2\sin\chi_3)(\sin\chi_1 \cos\chi_3+ \cos\chi_1\cos\chi_2\sin\chi_3)-\\\nonumber
&&
e^{i \lambda_3\tau} (\sin\chi_1 \sin\chi_3 - \cos\chi_1\cos\chi_2\cos\chi_3)(\cos\chi_1\sin\chi_3 + \sin\chi_1\cos\chi_2\cos\chi_3)\Big)|001\rangle +\\\nonumber
&&
\gamma_1(\tau)|010\rangle +\gamma_2(\tau) |100\rangle,
\end{eqnarray}
where we do not give explicite form for $\gamma_i$, $i=1,2$. 
Again, the probability of the excited state  transfer is defined by Eq.(\ref{theor0}).
\begin{figure*}[!]
\epsfig{file=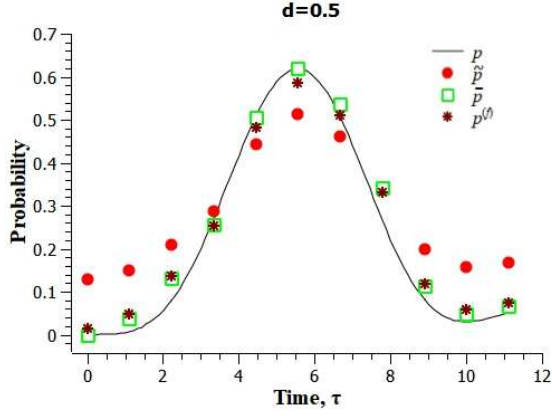,
  scale=0.7
   ,angle=0
}
\caption{ {Probability of the excited state transfer from the 1st to the 3rd spin  of the alternating spin chain, $d=0.5$,  governed by  the $XX$-Hamiltonian, interactions among all nodes are included.} {  
We present the theoretically calculated  probability  $p$ (solid lines), probability calculated   on the quantum processor $\tilde p$(circles),  corrected probabilities $\bar{p}$  (squares) and $p^{(f)}$  (stars)}. 
{To construct $\bar{p}$ we use 
 $s=0.1276$ and $k=1.6046$ in Eq.(\ref{sk}).}}
  \label{Fig:QE3} 
\end{figure*}

We emphasize that  the structure of the formulas for $U_{123}$ (\ref{3U123}) and $e^{-i \tilde \Lambda_{123} \tau}$ (\ref{3ev}) differ from the structure of the analogous  operators for the case of approximation of  nearest-neighbor approximations, see Eqs.(\ref{U123}) and (\ref{ev}).
{In fact,} the unitary operator $U_{123}$ in (\ref{3U123}) includes three operators of form 
(\ref{Uchi}) ($U_{12}$, $U_{23}$ and $U_{31}$) instead of two such operators in  Eq.(\ref{U123}), while  the form of the operator $e^{-i \tilde \Lambda_{123} \tau}$ coincides with the form of this operator in Eq.(\ref{Lambda2}).
{The scheme in Fig.\ref{Fig:S} does not fit this example. 
Nevertheless, the corrected probability $p^{(f)}$ with the parameters $\bar{s}$ and $\bar{k}$ from Table \ref{Table:1} (stars in Fig.\ref{Fig:QE3}) fits rather well the theoretical curve (solid line in Fig.\ref{Fig:QE3}), just demonstrating a certain universality of the proposed correction method. }



\section{Conclusion}
\label{Section:Conclusion}
Although the existing quantum processors are very noisy and usually give the results with large error, there are some methods to overcome this problem at least partially.  
One of such methods  is a two-parametric recovering  transformation combining properly adjusted shift and scale of  measured {probabilities} \cite{ZRPL}. We show that,
for a given scheme implemented on a quantum processor, there is such transformation  of the measured probability from $\tilde{p}$ to $p^{(f)}$ that the latter does not significantly differ from the theoretically predicted value $p$.
At that, the most important factor responsible for applicability of a recovery transformation with a particular values  of parameters is the number of CNOT operations included in the evolutionary operator under consideration. 

{We simulate  the evolution of the three-node spin chain governed by the XX-Hamiltonian on the quantum processor using  diagonalization of the Hamiltonian thus avoiding Trotterization.   The diagonalization allows to present  the time-dependent part of the evolution operator in the diagonal form.
 We represent both  the matrix of eigenvectors  and the (time-dependent) diagonal part of the evolution operator   in terms of CNOT operations and one-qubit rotations. We show that there is the recovering transformation  (\ref{barp}) with particular values of parameters $\bar{s}$ and $\bar{k}$ that improves results of calculations.
Having constructed the parameters  $\bar{s}$ and $\bar{k}$ for the alternating three-node chain governed by the $XX$-Hamiltonian under approximation of nearest-node interactions we then demonstrate its applicability to the case of the $XX$-Hamiltonian with all-node interactions just confirming certain universality of the considered correction method. This method is applicable if the measured probability $\gtrsim 0.2$. Otherwise the error still remains significant, as shown in the last plate of  Fig.\ref{Fig:QE1} ($d=0.2$).

Thus, we propose the scheme simulating three-qubit evolution governed by the $XX$-Hamiltonian and avoiding  Trotterization. This scheme is based on blocks $U_{ij}$ of form (\ref{Uchi}) having an important property of conserving the excitation number in a system. Extension of such scheme to systems of larger dimension would be of interest for simulating spin dynamics on a quantum processor.  
}


{\bf Acknowledgments} Authors acknowledge the use of the IBM Quantum Experience for this work. The viewpoints
expressed are those of the authors and do not reflect the official policy or position of IBM or the IBM
Quantum Experience team. We acknowledge funding from the Ministry of Science and Higher Education of the Russian Federation (Grant No. 075-15-2020-779).

{\bf Conflict of interest} The authors declare that they have no conflict of interest.

\end{document}